\documentclass[preprint]{elsarticle}

\usepackage{graphicx}
\usepackage{amsmath}
\usepackage{pifont}
\usepackage{geometry}
\usepackage{txfonts}
\usepackage{hyperref}
\usepackage{dcolumn}
\usepackage{bm}
\usepackage{multirow}

\begin{document}

\title{New discrete method for investigating the response properties in finite 
electric field}

\author[kimuniv]{Myong-Chol Pak\corref{cor}}
\author[kimuniv]{Nam-Hyok Kim}
\author[kimuniv]{Hak-Chol Pak}
\author[kimuniv]{Song-Jin Im}

\cortext[cor]{Corresponding author}

\address[kimuniv]{ Department of Physics, {\bf {\it Kim Il Sung}} university, Pyongyang , Democratic People's Republic of Korea}

\begin{abstract}

In this paper we develop a new discrete method for calculating the dielectric tensor and Born effective charge tensor in finite electric field by using Berry's phase and the gauge invariance. We present a new method to overcome non-periodicity of the potential in finite electric field due to the gauge invariance, and construct the dielectric tensor and Born effective charge tensor that satisfy translational symmetry in finite electric field. In order to demonstrate the correctness of this method, we also perform calculations for the semiconductors AlAs and GaAs under the finite electric field to compare with the preceding method and the experiment.
\end{abstract}

\maketitle

\section{\label{sec:md-intro}Introduction}
The investigation for calculating the dielectric tensor and Born effective charge tensor in finite 
electric field is very important in studying of bulk ferroelectrics, ferroelectric films, superlattices, 
lattice vibrations in polar crystals, and so on[1,2,3]. 

Recently, the investigation of response properties to the external electric field is becoming interested theoretically as well as practically. In particular, the dielectric tensor and Born effective 
charge tensor in finite electric field are important physical quantities for analyzing and modeling the response of material to the electric field. In case of zero electric field, these response properties have already been studied by using DFPT (Density Functional Perturbation theory), and excellent results have been obtained [1].

DFPT[4]provides a powerful tool for calculating the $2^{nd}$-order derivatives of the total energy 
of a periodic solid with respect to external perturbations, such as strains, atomic sublattice 
displacements, a homogeneous electric field etc. In contrast to the case of strains and sublattice 
displacements for which the perturbing potential remains periodic, treatment of homogeneous 
electric fields is subtle, because the corresponding potential requires a term that is linear in real 
space, thereby breaking the translational symmetry and violating the conditions of Bloch's theorem. Therefore, electric field perturbations have already been studied using the long-wave 
method, in which the linear potential caused by applied electric field is obtained by considering a 
sinusoidal potential in the limit that its wave vector goes to zero[5]. In this approach, however, the 
response tensor can be evaluated only at zero electric field.

In nonzero electric field, the investigation of the response properties can't be performed using 
method based on Bloch's theorem, for nonperiodicity of the potential with respect to electric field. Therefore, several methods for overcoming it have been developed [2,3].  

Ref.[2] introduces the electric field-dependent energy functional by Berry's phase, and suggests 
the methodology for calculating by using finite-difference scheme. Ref.[3] discusses the proposal for calculating it by the discretized form of Berry's phase term and response theory with respect to 
perturbation of the finite electric field. However, in these methods, the nonperiodicity of the 
potential due to electric field is resolved by introducing polarized WFs (Wannier Functions) due to 
finite electric field. This requires much cost in calculating its inverse matrix in the perturbation 
expansion of Berry's phase and yields instability of results. 

In this paper, we developed a new discrete method for calculating the dielectric tensor and Born 
effective charge tensor in finite electric field by using Berry's phase and the gauge invariance. We present a new method for overcoming non-periodicity of the potential in finite electric field due to the gauge invariance, and calculate the dielectric tensor and Born effective charge tensor in a discrete different way than ever before. 

This paper is organized as follows. In Sec. 2, instead of preceding investigation in which the 
total field-dependent energy functional is divided into Kohn-Sham energy, Berry's phase and 
Lagrange multiplier term, we discuss the method for studying the response properties with a new 
discrete way by using the polarization written with Berry's phase and unit cell periodic function 
polarized by field. In Sec. 3, we calculate the dielectric tensor and Born effective charge tensor in 
finite electric field by constructing the polarized Bloch wave Function and evaluating linear 
response of the wave function with Sternheimer equation. We also calculate the $2^{nd}$-order 
nonlinear dielectric tensor indicating nonlinear response property with respect to electric field. In 
order to demonstrate the correctness of the method, we also perform calculations for the 
semiconductors AlAs and GaAs under the finite electric field. In Sec. 4, summary and conclusion are presented.     

\section{\label{sec:md-method}New discrete method by using Berry's phase and the gauge invariance}
The response tensors with respect to electric field in finite electric field are presented by the $2^{nd}$-order derivatives of the field-dependent total energy functional with respect to the atomic sublattice displacements and the homogeneous electric field. Here, the field-dependent energy functional[6] is 
\begin{equation}
\label{eq:md-br1}
E[\{ u^{(\vec \varepsilon )} \} ,\vec \varepsilon ] = E_{KS} [\{ u^{(\vec \varepsilon )} \} ] - \Omega \vec \varepsilon  \cdot {\bf{P}}[\{ u^{(\vec \varepsilon )} \}
\end{equation}
where  $E_{KS},\vec \varepsilon,{\bf{P}}$  are the Kohn-Sham energy functional, the finite electric field, and the cell 
volume, respectively. In addition, $u^{(\vec \varepsilon )}$ is a set of unit cell periodic function polarized by field, and polarization ${\bf{P}}$  written through Berry's phase is 
\begin{equation}
\label{eq:md-br2}
{\bf{P}} =  - {{ife} \over {(2\pi )^3 }}\sum\limits_{n = 1}^M {\int\limits_{BZ} {d^3 k\left\langle {u_{n{\bf{k}}}^{(\vec \varepsilon )} } \right|} } \nabla _{\bf{k}} \left| {u_{n{\bf{k}}}^{(\vec \varepsilon )} } \right\rangle
\end{equation}
where $f$ is the spin degeneracy, and $f = 2$. 

In fact, the polarization is calculated by the following discretized form that is suggested by King-Smith and Vanderbilt[7]. 
\begin{equation}
\label{eq:md-br3}
{\bf{P}} = {{ef} \over {2\pi \Omega }}\sum\limits_{i = 1}^3 {{{{\bf{a}}_i } \over {N_ \bot ^{(i)} }}} \sum\limits_{l = 1}^{N_ \bot ^{(i)} } {{\mathop{\rm Im}\nolimits} \ln \prod\limits_{j = 1}^{N_i } {\det S_{{\bf{k}}_{lj} ,{\bf{k}}_{lj + 1} } } },
\end{equation}
(Ref.[6] and [7] point out the meaning of every parameter in Eq.~\ref{eq:md-br2} and Eq.~\ref{eq:md-br3}.) Next, if we consider the orthonormality constraints of the unit cell periodic function polarized by field
\begin{equation}
\label{eq:md-br4}
\left\langle {{u_{m{\bf{k}}}^{(\vec \varepsilon )} }}
 \mathrel{\left | {\vphantom {{u_{m{\bf{k}}}^{(\vec \varepsilon )} } {u_{n{\bf{k}}}^{(\vec \varepsilon )} }}}
 \right. \kern-\nulldelimiterspace}
 {{u_{n{\bf{k}}}^{(\vec \varepsilon )} }} \right\rangle  = \delta _{mn}
\end{equation}
,the total energy functional is divided into 3 parts as follows. 
\begin{equation}
\label{eq:md-br5}
F = F_{KS}  + F_{BP}  + F_{LM} 
\end{equation}
where $F_{KS}  = E_{KS}$ is Kohn-Sham energy and $F_{BP}  =  - \Omega \vec \varepsilon  \cdot {\bf{P}}$ is the coupling between the electric field and the polarization by Berry's phase, and the constraints are given by Lagrange multiplier term, $F_{LM}$. Next, the set of unit cell periodic functions polarized by field, $\{ u^{(\vec \varepsilon )} \}$ is determined with variational method. The set of its function is different from a set of unit cell periodic function in zero field. Although, strictly speaking, calculated ground state is not exact ground state, this method is a way to overcome nonperiodicity of the potential caused by electric field[3]. Therefore, it does not include explicitly the gauge invariant property and requires big cost in calculating its inverse matrix in the perturbation expansion of Berry's phase, yielding instability of results.

We apply the perturbation expansion by using DFPT, and investigate the response property with a new discrete method by using Eq.~\ref{eq:md-br2} and unit cell periodic functions polarized by field. Since the general perturbation expansion methods were mentioned in Refs.[1,2,3], we consider the response tensors, dielectric tensor and Born effective charge tensor in case of perturbation with respect to the atomic sublattice displacements and the homogeneous electric field. 

In Gaussian system the dielectric tensor is
\begin{equation}
\label{eq:md-br6}
\in _{\alpha \beta }  = \delta _{\alpha \beta }  + 4\pi \chi _{\alpha \beta } 
\end{equation}
and then electric susceptibility tensor can be written by perturbation expansion.
\begin{equation}
\label{eq:md-br7}
\begin{split}
\chi _{\alpha \beta } &=  - {1 \over \Omega }{{\partial ^2 F} \over {\partial \varepsilon _\alpha  \partial \varepsilon _\beta  }} =  - {f \over {2(2\pi )^3 }}\int\limits_{BZ} {d^3 k\sum\limits_{n = 1}^M {[\left\langle {u_{n{\bf{k}}}^{\varepsilon _\alpha  } } \right|T + v_{ext} \left| {u_{n{\bf{k}}}^{\varepsilon _\beta  } } \right\rangle  + } } \left\langle {u_{n{\bf{k}}}^{\varepsilon _\beta  } } \right|T + v_{ext} \left| {u_{n{\bf{k}}}^{\varepsilon _\alpha  } } \right\rangle ]\\
&+ \sum\limits_{n = 1}^M {[\left\langle {u_{n{\bf{k}}}^{\varepsilon _\alpha  } } \right|(ie{\partial  \over {\partial k_\beta  }})\left| {u_{n{\bf{k}}}^{(0)} } \right\rangle  + \left\langle {u_{n{\bf{k}}}^{(0)} } \right|(ie{\partial  \over {\partial k_\beta  }})\left| {u_{n{\bf{k}}}^{\varepsilon _\alpha  } } \right\rangle  + \left\langle {u_{n{\bf{k}}}^{\varepsilon _\beta  } } \right|(ie{\partial  \over {\partial k_\alpha  }})\left| {u_{n{\bf{k}}}^{(0)} } \right\rangle }\\
&+ \left\langle {u_{n{\bf{k}}}^{(0)} } \right|(ie{\partial  \over {\partial k_\alpha  }})\left| {u_{n{\bf{k}}}^{\varepsilon _\beta  } } \right\rangle ] + {f \over {2(2\pi )^3 }}\int\limits_{BZ} {d^3 k\sum\limits_{m,n = 1}^M {\Lambda _{mn}^{(0)} ({\bf{k}})[\left\langle {{u_{n{\bf{k}}}^{\varepsilon _\alpha  } }}
 \mathrel{\left | {\vphantom {{u_{n{\bf{k}}}^{\varepsilon _\alpha  } } {u_{m{\bf{k}}}^{\varepsilon _\beta  } }}}
 \right. \kern-\nulldelimiterspace}
 {{u_{m{\bf{k}}}^{\varepsilon _\beta  } }} \right\rangle  + \left\langle {{u_{n{\bf{k}}}^{\varepsilon _\beta  } }}
 \mathrel{\left | {\vphantom {{u_{n{\bf{k}}}^{\varepsilon _\beta  } } {u_{m{\bf{k}}}^{\varepsilon _\alpha  } }}}
 \right. \kern-\nulldelimiterspace}
 {{u_{m{\bf{k}}}^{\varepsilon _\alpha  } }} \right\rangle ]} }  - {1 \over {2\Omega }}{{\partial ^2 E_{XC} } \over {\partial \varepsilon _\alpha  \partial \varepsilon _\beta  }}
\end{split}
\end{equation}
Though Eq.~\ref{eq:md-br7} reflects successfully the response properties with respect to perturbation in finite electric field, it does not describe sufficiently the periodic effect of crystal. Because the operator,$ie\nabla _{\bf{k}}$ hidden Berry's phase must be applied to gauge invariant quantity in order to overcome nonperiodicity of potential caused by field[7]. Therefore, using the gauge invariant form,$ie{\partial  \over {\partial k_\alpha  }}\sum\limits_{m = 1}^M {\left| {u_{m{\bf{k}}}^{(0)} } \right\rangle } \left\langle {u_{m{\bf{k}}}^{(0)} } \right|$ and considering $0^{th}$-order,$\Lambda _{mn}^{(0)} ({\bf{k}}) = \varepsilon _{n{\bf{k}}}^{(0)} \delta _{mn}$, dielectric tensor is  
\begin{equation}
\label{eq:md-br8}
\begin{split}
\chi _{\alpha \beta } &= \left. { - {1 \over \Omega }{{\partial ^2 F} \over {\partial \varepsilon _\alpha  \partial \varepsilon _\beta  }}} \right|_{\varepsilon  = \varepsilon ^{(0)} }\\
& =  - {f \over {2(2\pi )^3 }}\int\limits_{BZ} {d^3 k\sum\limits_{n = 1}^M {[\left\langle {u_{n{\bf{k}}}^{\varepsilon _\alpha  } } \right|T + v_{ext}  - \varepsilon _{n{\bf{k}}}^{(0)} \left| {u_{n{\bf{k}}}^{\varepsilon _\beta  } } \right\rangle  + } } \left\langle {u_{n{\bf{k}}}^{\varepsilon _\beta  } } \right|T + v_{ext}  - \varepsilon _{n{\bf{k}}}^{(0)} \left| {u_{n{\bf{k}}}^{\varepsilon _\alpha  } } \right\rangle\\
& + \left\langle {u_{n{\bf{k}}}^{\varepsilon _\alpha  } } \right|(ie{\partial  \over {\partial k_\beta  }}\sum\limits_{m = 1}^M {\left| {u_{m{\bf{k}}}^{(0)} } \right\rangle } \left\langle {u_{m{\bf{k}}}^{(0)} } \right|)\left| {u_{n{\bf{k}}}^{(0)} } \right\rangle  + \left\langle {u_{n{\bf{k}}}^{\varepsilon _\beta  } } \right|(ie{\partial  \over {\partial k_\alpha  }}\sum\limits_{m = 1}^M {\left| {u_{m{\bf{k}}}^{(0)} } \right\rangle } \left\langle {u_{m{\bf{k}}}^{(0)} } \right|)\left| {u_{n{\bf{k}}}^{(0)} } \right\rangle\\
&- \left\langle {u_{n{\bf{k}}}^{(0)} } \right|(ie{\partial  \over {\partial k_\beta  }}\sum\limits_{m = 1}^M {\left| {u_{m{\bf{k}}}^{(0)} } \right\rangle } \left\langle {u_{m{\bf{k}}}^{(0)} } \right|)\left| {u_{n{\bf{k}}}^{\varepsilon _\alpha  } } \right\rangle  - \left\langle {u_{n{\bf{k}}}^{(0)} } \right|(ie{\partial  \over {\partial k_\alpha  }}\sum\limits_{m = 1}^M {\left| {u_{m{\bf{k}}}^{(0)} } \right\rangle } \left\langle {u_{m{\bf{k}}}^{(0)} } \right|)\left| {u_{n{\bf{k}}}^{\varepsilon _\beta  } } \right\rangle ] - \left. {{1 \over {2\Omega }}{{\partial ^2 E_{XC} } \over {\partial \varepsilon _\alpha  \partial \varepsilon _\beta  }}} \right|_{\varepsilon  = \varepsilon ^{(0)} }
\end{split}
\end{equation}
where BZ(Brillouin Zone) integration is performed by Monkhorst-Pack special point method. Meanwhile, the partial derivative is calculated with following discretized method. 
\begin{equation}
\label{eq:md-br9}
{\partial  \over {\partial k_x }}\left| {u_{m,i,j,k}^{} } \right\rangle \left\langle {u_{m,i,j,k}^{} } \right| = {1 \over {2\Delta k_x }}(\left| {u_{m,i + 1,j,k}^{} } \right\rangle \left\langle {u_{m,i + 1,j,k}^{} } \right| - \left| {u_{m,i - 1,j,k}^{} } \right\rangle \left\langle {u_{m,i - 1,j,k}^{} } \right|)
\end{equation}
\begin{equation}
\label{eq:md-br10}
{\partial  \over {\partial k_y }}\left| {u_{m,i,j,k}^{} } \right\rangle \left\langle {u_{m,i,j,k}^{} } \right| = {1 \over {2\Delta k_y }}(\left| {u_{m,i,j + 1,k}^{} } \right\rangle \left\langle {u_{m,i,j + 1,k}^{} } \right| - \left| {u_{m,i,j - 1,k}^{} } \right\rangle \left\langle {u_{m,i,j - 1,k}^{} } \right|)
\end{equation}
\begin{equation}
\label{eq:md-br11}
{\partial  \over {\partial k_z }}\left| {u_{m,i,j,k}^{} } \right\rangle \left\langle {u_{m,i,j,k}^{} } \right| = {1 \over {2\Delta k_z }}(\left| {u_{m,i,j,k + 1}^{} } \right\rangle \left\langle {u_{m,i,j,k + 1}^{} } \right| - \left| {u_{m,i,j,k - 1}^{} } \right\rangle \left\langle {u_{m,i,j,k - 1}^{} } \right|)
\end{equation}
Additionally, the $1^{st}$-order wave function response with respect to finite electric field is calculated with the following Sternheimer equation
\begin{equation}
\label{eq:md-br12}
P_{c{\bf{k}}} (T + v_{ext}  - \varepsilon _{n{\bf{k}}}^{(0)} )P_{c{\bf{k}}} \left| {u_{n{\bf{k}}}^{\varepsilon _\alpha  } } \right\rangle  =  - P_{c{\bf{k}}} (ie{\partial  \over {\partial k_\alpha  }}\sum\limits_{m = 1}^M {\left| {u_{m{\bf{k}}}^{(0)} } \right\rangle } \left\langle {u_{m{\bf{k}}}^{(0)} } \right|)\left| {u_{n{\bf{k}}}^{(0)} } \right\rangle
\end{equation}
Generally, investigation of the $2^{nd}$-order energy response requires up to $1^{st}$-order wave function response with respect to perturbation by using $"$2n+1 $"$theorem. Therefore, every result can be calculated with only the $1^{st}$-order wave function response to finite electric field.

In this way, Born effective charge tensor is 
\begin{equation}
\label{eq:md-br13}
\begin{split}
Z_{\kappa ,\alpha \beta }^* &= \left. { - {{\partial ^2 F} \over {\partial \varepsilon _\alpha  \partial \tau _{\kappa , \beta }}}} \right|_{\varepsilon  = \varepsilon ^{(0)} }\\
& = {{f\Omega } \over {(2\pi )^3 }}\int\limits_{BZ} {d^3 k\sum\limits_{n = 1}^M {[\left\langle {u_{n{\bf{k}}}^{(0)} } \right|(T + v_{ext} )^{\tau _{\kappa ,\beta } } \left| {u_{n{\bf{k}}}^{\varepsilon _\alpha  } } \right\rangle  + \left\langle {u_{n{\bf{k}}}^{\varepsilon _\alpha  } } \right|(T + v_{ext} )^{\tau _{\kappa ,\beta } } \left| {u_{n{\bf{k}}}^{(0)} } \right\rangle  + \left. {{{\partial ^2 E_{XC} } \over {\partial \tau _{\kappa ,\beta}  \partial \varepsilon _\alpha  }}} \right|_{\varepsilon  = \varepsilon ^{(0)} } } } 
\end{split}
\end{equation}
Eq.~\ref{eq:md-br13} also calculate with DFPT and the wave function polarized by field.  

\section{\label{sec:md-result}Results and Analysis}

The calculation of the dielectric permittivity tensor and the Born effective charge tensor is performed in three steps. First, a ground state calculation in finite electric field is performed using the Berry's phase method implemented in the ABINIT code, and the field-polarized Bloch functions are stored for the later linear-response calculation. Second, the linear-response calculation is performed to obtain the first order response of Bloch functions. Third, the matrix elements of the dielectric and Born effective charge tensors are computed using these $1^{st}$-order responses.

To verify the correctness of our method, we have performed test calculation on two 
prototypical semiconductors, AlAs and GaAs. In this calculation, we have used the HSC norm-conserving pseudopotential method based on Density Functional Theory with LDA (Local Density Approximation). The cutting energy, $E_{cut}  = 20Ry$ and $6 \times 6 \times 6$  Monkhorst-Pack mesh for $k$-point sampling were used.

In Table 1, we present the calculated values of dielectric tensor and Born effective charge tensor of AlAs and GaAs, when such finite electric field as in Ref. [3] is applied along the [100] direction. In order to compare our method with the preceding one, we present the calculated values in our method and preceding one (Ref.[3]), and the experimental values. As you see in Table 1, the calculated value of dielectric tensor in our method goes to the experiment one[2] more closely than the calculated one in preceding method (Ref.[3]).However, in case of Born effective charge tensor, the difference between our method and preceding one does not almost occur. It shows that in calculating Born effective charge tensor, there exist the $1^{st}$-order contribution of the potential with respect to atomic sublattice displacements and one of the polarized wave function with respect to the finite electric field, the latter playing the essential role.  
\begin{table*}[!h]
\begin{center}
\caption{\label{tab:md-die}Calculated and experimental values of dielectric tensor and Born effective charge tensor in finite electric field}
\begin{tabular}{|c|c|c|c|}
\hline
Material& Method & $\in$ & $Z^*$ \\
\hline
 & Our Method & 9.48 & 2.05 \\
AlAs &Preceding Method[3] & 9.72 & 2.03 \\
 & Experiment[2] & 8.2 & 2.18 \\
\hline
 & Our Method & 12.56 & 2.20 \\
GaAs & Preceding Method[3] & 13.32 & 2.18 \\
 & Experiment[2] & 10.9 & 2.07 \\
\hline
\end{tabular}
\end{center}
\end{table*}

We also calculated the $2^{nd}-order \ nonlinear \ dielectric \ tensor$, nonlinear response property with respect to electric field. The $2^{nd}$-order nonlinear dielectric tensor is
\begin{equation}
\label{eq:md-br14}
\chi _{123}^{(2)}  = {1 \over 2}{{\partial ^2 P_2 } \over {\partial \varepsilon _1 \partial \varepsilon _3 }} = {1 \over 2}{{\partial \chi _{23} } \over {\partial \varepsilon _1 }}
\end{equation}
Table 2 shows calculated value of the $2^{nd}$-order nonlinear dielectric tensor on AlAs.
\begin{table*}[!h]
\begin{center}
\caption{\label{tab:md-non}Calculated value of the $2^{nd}$-order nonlinear dielectric tensor on AlAs}
\begin{tabular}{|c|c|}
\hline
Method & $\chi _{123}^{(2)} (pm/V)$ \\
\hline
Our Method & 67.32 \\
Preceding Method[3] & 60.05 \\
Experiment[8] & 78 $\pm$ 20\\
\hline
\end{tabular}
\end{center}
\end{table*}
As shown in Table 2, the calculated value of $2^{nd}$-order nonlinear dielectric tensor in our method coincides with the experimental value[8] more closely than the calculated one in the preceding method (Ref.[3]).

\section{\label{sec:sum}Summary}

We suggested a new method for calculating the  dielectric tensor and Born effective charge tensor in finite electric field. In particular, in order to overcome nonperiodicity of potential caused by electric field, a new transformation conserving gauge invariant property is introduced. In future, this methodology can be expanded not only to perturbation with respect to field and atomic replacement but also to the other cases, such as strain and chemical composition of solid solution.

\section*{\label{ack}Acknowledgments}

It is pleasure to thank Jin-U Kang, Chol-Jun Yu, Kum-Song Song, Kuk-Chol Ri and 
Song-Bok Kim for useful discussions. This work was supported by the Physics faculty in {\bf Kim Il Sung} university of Democratic People's Republic of Korea.    

\section*{\label{ref}References}

[1] C.-J. Yu and H. Emmerich. J. Phys.:Condens. Matter, 19:306203, 2007.

[2] I. Souza, J. Iniguez, and D. Vanderbilt. Phys. Rev. Lett., 89:117602, 2002.

[3] X. Wang and D. Vanderbilt. Phys. Rev. B, 75:115116, 2007.

[4] S. Baroni, Stefano de Gironcoli, and Andrea Dal Corso. Rev.Mod.Phys., 73:515, 2001.

[5] X. Gonze and C. Lee. Phys. Rev. B, 55:10355, 1997.

[6] R. W. Nunes and X. Gonze. Phys. Rev. B, 63:155107, 2001.

[7] R. D. King-Smith and D.Vanderbilt. Phys. Rev. B, 48:4442, 1993.

[8] I. Shoji, T. Kondo, and R. Ito. Opt. Quantum Electron, 34:797, 2002

\end{document}